\documentclass[pdflatex,sn-mathphys-num]{sn-jnl}
\usepackage{graphicx}%
\usepackage{multirow}%
\usepackage{amsmath,amssymb,amsfonts}%
\usepackage{amsthm}%
\usepackage{mathrsfs}%
\usepackage[title]{appendix}%
\usepackage{xcolor}%
\usepackage{textcomp}%
\usepackage{booktabs}%
\usepackage{subcaption}%
\usepackage{algorithm}%
\usepackage{algorithmicx}%
\usepackage{algpseudocode}%
\usepackage{listings}%

\theoremstyle{thmstyleone}%
\theoremstyle{thmstyletwo}%

\theoremstyle{thmstylethree}%

\makeatother
\begin{document}

\title[Article Title]{Crystal Structure Based Graph Neural Network And LLM Multi-Modal Approach for Material Magnetic Ordering Prediction}

\author*[1]{\fnm{Adam} \sur{Zheng}}\email{adam.ray.zheng@gmail.com}

\author[2]{\fnm{April} \sur{Zheng}}\email{aprilzheng@college.harvard.edu}

\author[3]{\fnm{Sheng} \sur{Ran}}\email{rans@wustl.edu}

\affil[1]{\orgaddress{\street{1755 Merivale Rd}, \city{Nepean}, \postcode{K2G 1E}, \state{Ontario}, \country{Canada}}}

\affil[2]{\orgname{Harvard University}, \city{Cambridge}, \postcode{02138}, \state{Massachusetts}, \country{USA}}

\affil[3]{\orgdiv{Department of Physics}, \orgname{Washington University in St. Louis}, \orgaddress{\street{516 Main Street}, \city{St. Louis}, \postcode{MO 63130-4899}, \state{Missouri}, \country{USA}}}

\abstract{  AI/ML models are useful for the prediction of materials' properties. Recently, there has been great progress in developing Neural Network models to predict the properties of solids, molecules, and crystals, avoiding expensive and time-consuming laboratory testing. This study presents a novel AI/ML-based multi-modal approach to predict material magnetic ordering using material crystal structure information as input to a Graph Neural Network model (GNN) and semantic text descriptions of the material's properties and structure as input to a Large Language Model (LLM), before performing information fusion of the GNN and LLM outputs to predict material magnetic ordering in four classes (Non-magnetic, Ferromagnetic, Antiferromagnetic, Ferrimagnetic).  The model is trained, validated, and tested on a dataset which is a combination of Materials Project dataset, Magndata dataset, and a set of uranium compounds with magnetic ordering determined in laboratory testing, with an 80/10/10 (training/validation/testing) data split. The proposed GNN and LLM multi-modal approach achieves better overall accuracy than both the standalone GNN and LLM approaches with three fusion strategies.}


\maketitle

\section{Introduction}
\label{introduction}
Material properties are critical to material engineering and determine where materials can be applied. For example, the energy band gap can be used to determine electricity conductivity and optical properties, the formation energy can be used to determine material stability, and magnetic order can indicate material performance under different thermal, mechanical, or electrical conditions. This information is then used to determine if the material can be used for different applications such as shielding, sensors, and actuators. 

Experimental determination of material properties, while accurate, can be very time-consuming and costly. To mitigate this problem, Density Functional Theory (DFT) is proposed in \cite{dft1,dft2,dft3} as a way to mathematically model the quantum mechanical behavior of atoms and electrons, which can be used to mathematically predict material properties. Though this mathematical approach is very popular, its prediction accuracy is heavily dependent on known approximation functions, and it cannot simulate complicated compound structure accurately. 

Recent research has focused on the integration of AI/ML based approaches into this field, and there has been much progress in developing Graph Neural Network models to accurately predict properties of solids, molecules and crystals, avoiding traditional laboratory testing. 
\citet{NIPS2017_303ed4c6} created SchNet, one of the first influential GNNs capable of modeling local interactions in the molecule without requiring atomic positions to lie on a fixed grid (a challenge faced by convolution neural networks (CNN)). These interactions were captured using interaction blocks that aggregated information from neighboring atoms and would be weighted atom-wise to produce a final prediction. \citet{PhysRevLett.120.145301} introduced the Crystal Graph Convolutional Neural Network (CGCNN) which adapted the previous framework onto crystalline compounds, encoding atoms as nodes and bonds as edges in a crystal graph and producing predictions that could match the accuracy of DFT-computed properties across a broad range of crystal types. \citet{D0CP01474E} introduced the Global Attention Graph Neural Network (GATGNN) model, which differentiates the contributions of individual atoms further by introducing a global attention layer and augmenting graph-attention layers and incorporates temperature, pressure, and entropy as global state inputs. In 2021, \citet{Choudhary} introduced the Atomistic Line Graph Neural Network (ALIGNN) model, one of the first models to effectively incorporate bond angle data as input, which achieved top prediction precision at time of proposal. 

Since the introduction of transformer architecture, the huge success of transformer-based large language models (LLMs) proposed by \citet{NIPS2017_3f5ee243} to process language-based problems demonstrates the potential of utilizing a transformer-based approach to tackle problems in the domain of material properties prediction. \citet{yan2024complete} proposed a transformer designed specifically for crystalline materials, employing invariant geometric descriptors of Euclidean distances and angles and equivariant vector representation. \citet{walker2021impact} proposed MatBERT (Bidirectional Encoder Representations from Transformers (BERT)), which uses contextualized word representations and semantically rich sentence embeddings for material prediction. \citet{rubungo2023llmproppredictingphysicalelectronic} proposed LLM-Prop, which leverages the general purpose learning capabilities of large language models (LLMs) to predict properties of crystals from their text descriptions and outperforms MatBERT. \citet{li2025hybrid} proposed a hybrid approach to use pretrained GNN and LLM models to encode material information including crystal structure and a tree-based ML regressor to perform property prediction.

In this paper, we present a Graph Neural Network (GNN) and LLM multi-modal approach based on work in \cite{li2025hybrid} to perform magnetic ordering prediction on material compounds as studied in \citet{PhysRevMaterials.8.114405}. Instead of using a pretrained GNN and LLM as information embedding only, a multi-modal approach is used to design the model. The model includes four blocks: GNN, LLM, information fusion, and classification, and all four blocks will be trained together. Three information fusion approaches are tested to show that information fusion provides slightly better accuracy.

\section{Methods}
\label{gen_inst}
\subsection{Model architecture}
\begin{figure}
    \centering
    \includegraphics[width=1\textwidth]{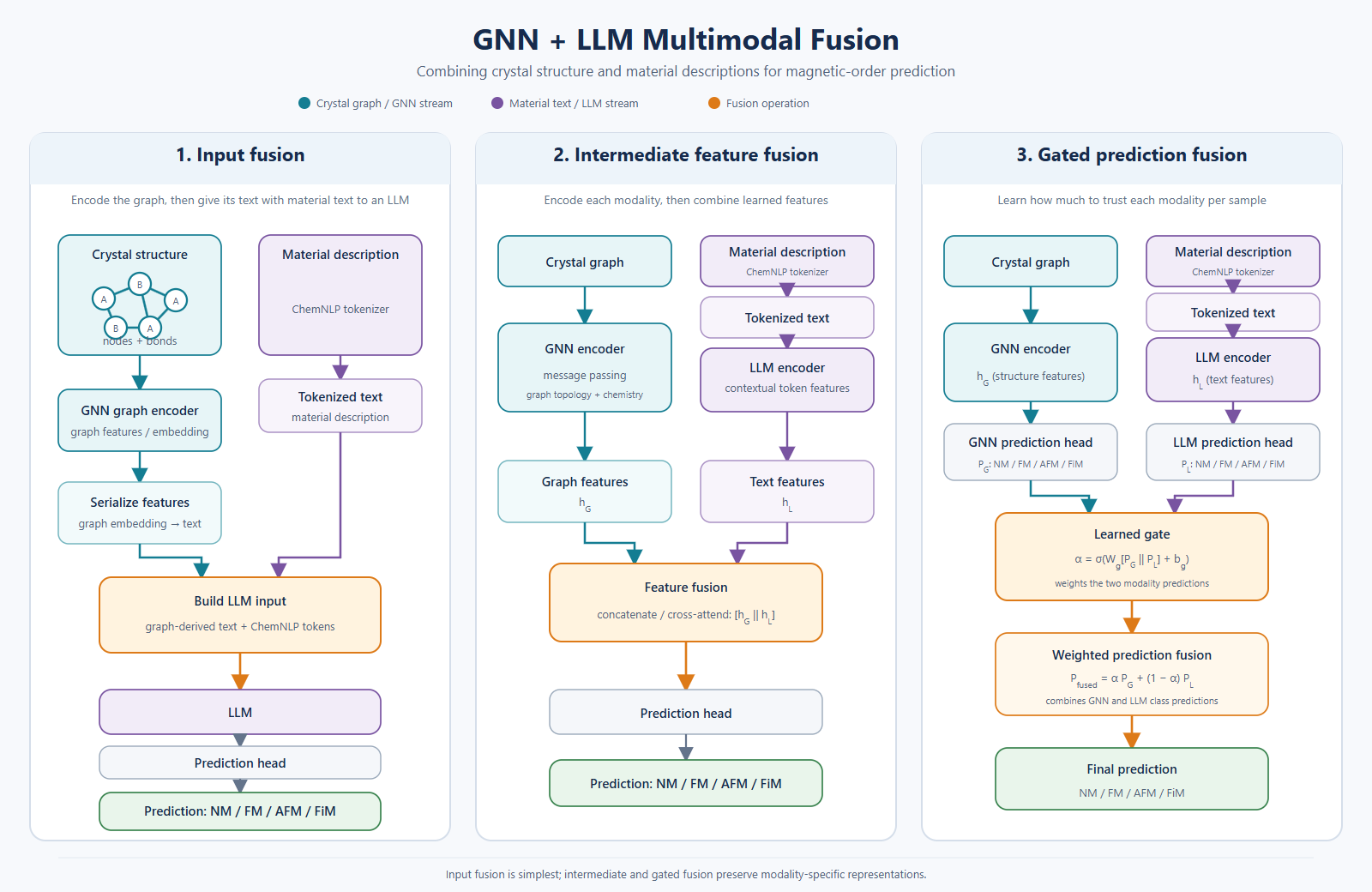}
    \caption{Architecture of ALIGNN-LLM multi-modal classifier}
    \label{fig:arch_mm}
\end{figure}

The GNN-LLM multi-modal classifier combines crystal structure analysis via a GNN with semantic text descriptor analysis via an LLM for predicting magnetic ordering (NM/FM/AFM/FiM). The GNN used in this approach is ALIGNN, and the LLM used is GPT-2 \cite{gpt2}. 
The model architecture is shown in Fig.~\ref{fig:arch_mm}. In this study, 3 fusion strategies are  evaluated: 
\begin{enumerate}
 \item Early fusion: fuse serialized GNN graph and tokenized material descriptors together as input to LLM for feature computation, and feed to classification head for final prediction
 \item Intermediate fusion: fuse GNN and LLM output features through concatenation and feed to classification head to compute the final class prediction
 \item Gated late fusion: create a learnable gate for each class that takes in GNN and LLM output logits and balances the weights for both branches for final prediction of classes 
\end{enumerate}

\subsection{Blocks}
\begin{figure}
    \centering
    \includegraphics[width=1\textwidth]{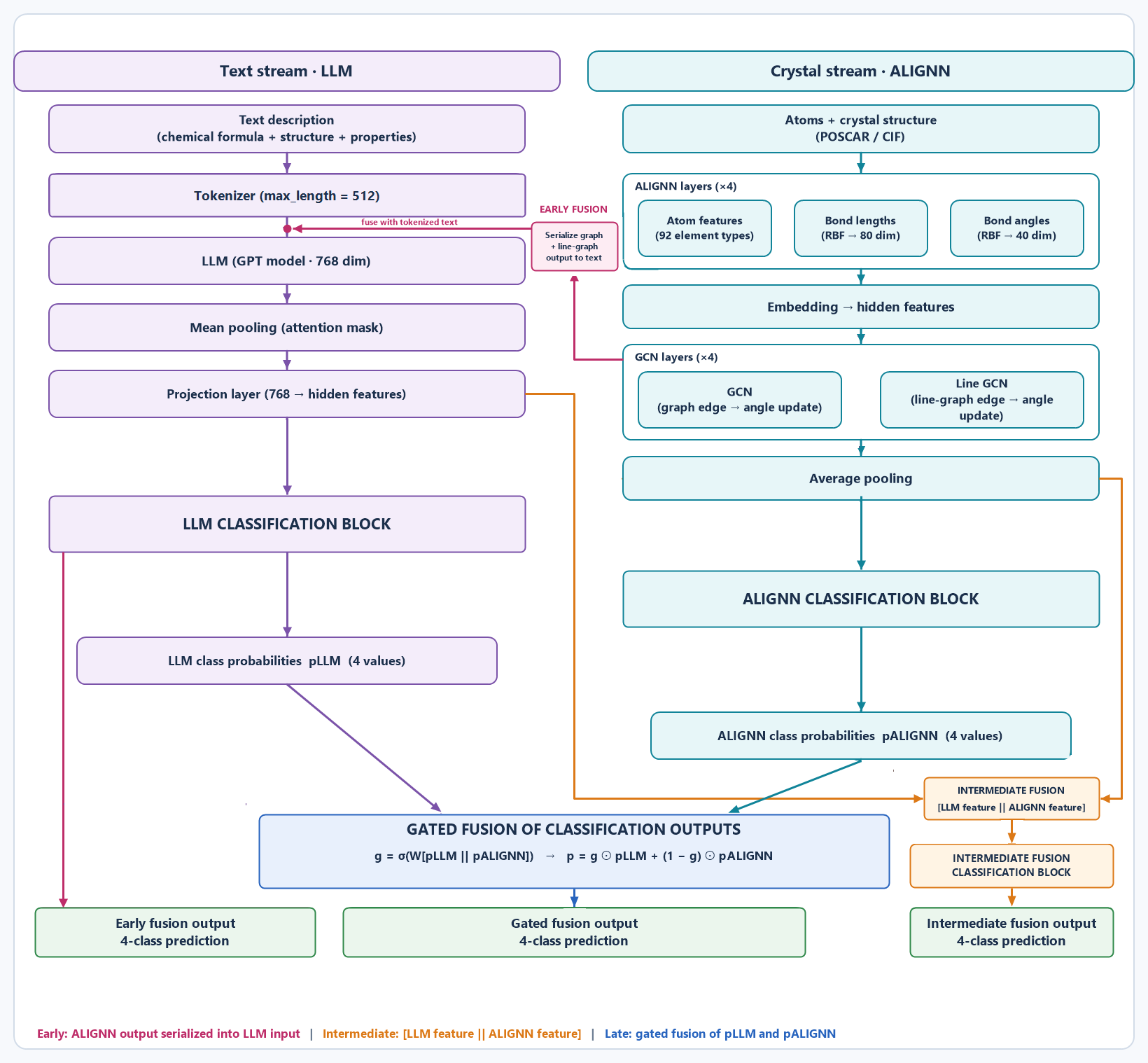}
    \caption{Architecture of ALIGNN-LLM multi-modal classifier}
    \label{fig:blocks_mm}
\end{figure}

The implementation blocks are illustrated in Fig.~\ref{fig:blocks_mm} including GNN (ALIGNN), LLM (GPT-2) and classification head. The workflow and fusion location of the three fusion approaches are illustrated.

\subsubsection{GNN Block (ALIGNN)}
Material structure information is first extracted in Pymatgen Structure JSON format and then encoded using Jarvis Atom functionality \cite{javisatoms}. Then, it is converted to a graph and line graph (which is the required input format for ALIGNN) via edge and angle embedding using RBF expansion and MLP (Multi-Layer Perceptron). ALIGNN includes the following two graphs: a Graph Convolutional Network (GCN) using atoms as nodes and bonds as edges, and a line GCN using bonds as nodes and bond angle triplets as edges. In this manner, more structural information can be captured and used to predict more accurately.

The following example illustrates the dimensions of the ALIGNN GCN and line GCN using data from MP-149 (silicon) from the Materials Project database \cite{mproj}:

For the GCN, atoms serve as the nodes and atomic bonds are the edges. CGCNN (Crystal Graph Convolutional Neural Network) embedding is used to represent each atom, and provides a 92-dimensional feature vector for each atom based on its properties from the periodic table. There is one node per Si atom; thus, the atom features have dimensions [2, 92]. Each atom has 28 neighbors within a cutoff of 5Å, and for each neighbor there is one directed ingoing and outgoing edge; hence, there are 56 directed bonds for each atom, and 56 edges per node for the GCN.

For the line GCN, the atom bonds act as the nodes and angle triplets act as the edges. To determine the number of angle triplets, each atom has 28 incoming and outgoing bonds, and thus the number of bonds including each atom is 784. Hence, for each bond (which includes two atoms), the number of angle triplets is $784\cdot 2=1568$; the total is 1544 after removing duplicates for both atoms.

\begin{figure}[h]
    \centering
    \includegraphics[width=0.75\textwidth]{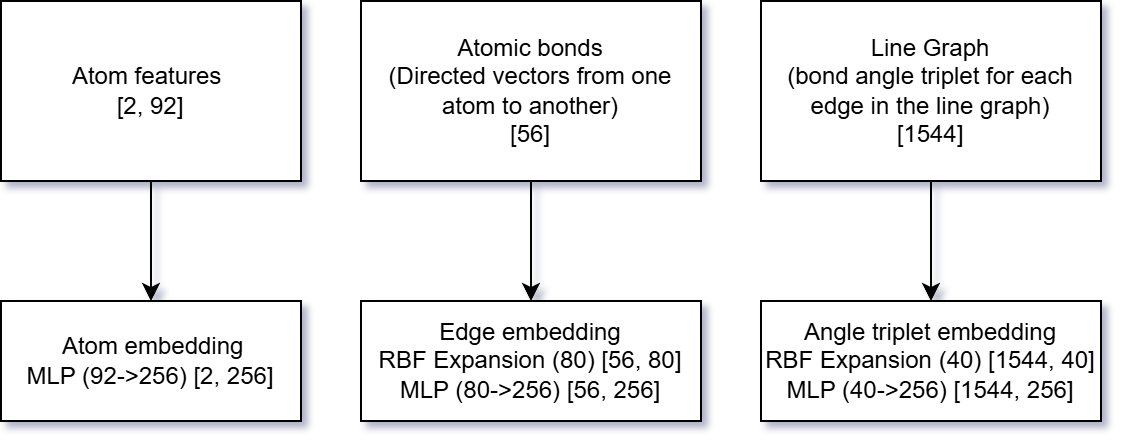}
    \caption{ALIGNN Graph construction}
    \label{fig:atom_edge}
\end{figure}

The graph construction (including data dimension changes) is demonstrated in Fig.~\ref{fig:atom_edge}.

\subsubsection{LLM Block}
The Material compound is first converted to a text description based on the ChemNLP implementation in \cite{chemnlp}. As an example, the information of MP-149 from Materials Project is converted to the following text description:

"The chemical information includes: The chemical has an atomic formula of Si with a prototype of A; Its molecular weight is 28.09 g/mol; The atomic fractions are Si: 1.0, and the atomic values X and Z are 1.9 and 14, respectively. The structure information includes: The lattice parameters are 3.85, 3.85, 3.85 with angles 60.0, 60.0, 60.0 degrees; The top K XRD peaks are found at 28.4, 47.2, 56.0, 69.0, 76.2 degrees; The material has a density of 2.313 g/cm³; The bond distances are as follows: Si-Si: 2.36, 3.85."

In this study, GPT-2 model is used in the LLM block. During testing, the base GPT-2 model with 124.64M parameters was found to be capable of handling the size of the dataset and complexity of the classification problem.

\subsubsection{Fusion }

The three fusion workflows are shown in Fig.~\ref{fig:blocks_mm} and the data fusion happens at input, feature and classification logits respectively. 

\subsubsection{Classification Block}

The classification block takes the fused information from the information fusion block and performs magnet ordering prediction using MLP based classification head.

\section{Data preparation and training}
\label{headings}

\subsection{Data Preparation}

A preliminary dataset consisting of 210,579 compounds was extracted from 3 sources: 
database.xlsx from \citet{PhysRevMaterials.8.114405}, Materials Project from \cite{mproj}, and Magndata from \cite{magndata}. database.xlsx includes 1064 compounds, Magndata includes 231 compounds, and MP includes 209,284 compounds. When there is duplication on certain material compounds in different sources, the ones with mp ordering being laboratory validated are selected over material project ones.

Compounds without an indicated
magnetic ordering (i.e. non-magnetic, ferromagnetic, ferrimagnetic, or
antiferromagnetic) were filtered out. The remaining dataset contained 200,431 compounds, with the following distribution of magnetic orderings as shown in Table~\ref{data-table}:


\begin{table}[!ht]
  \caption{Distribution of data in final dataset}
  \label{data-table}
  \centering
  \begin{tabular}{lll}
    \toprule
    Magnetic Ordering      & Number of data entries & Proportion of complete dataset      \\
    \midrule
    All (across all 4 classes)  & 200,431  & 100\%     \\
    \midrule
    Non-magnetic     & 119,225 & 59.48\%       \\
    \midrule
    Ferromagnetic     & 64,253  & 32.06\%       \\
    \midrule
    Antiferromagnetic     & 3,709   & 1.85\%       \\
    \midrule
    Ferrimagnetic     & 13,244  & 6.61\%       \\
    \bottomrule
  \end{tabular}
\end{table}

As seen in Table~\ref{data-table}, the distribution of data is unequal - there is significantly more data for the non-magnetic class than the others. For classification problems, optimal performance of AI/ML approaches requires a relatively balanced proportion of classes in the dataset. This poses a big challenge and the multi-modal approach can provide improved accuracy as shown in Section \ref {test_result}.  

\subsection{Training Procedure}
\label{traing_pro}

The dataset is split into training, validation, and testing sets with 80\%, 10\%, and 10\% of total dataset used for each respectively. During training, the training and validation accuracy is evaluated at the end of each epoch and the model with the highest validation accuracy up to that epoch is saved. After training, the best model is tested on the testing dataset to determine its final classification accuracy. The training/validation/testing are performed on DUAL Nvidia 3090 GTX GPU.

The three models are trained for 30 epochs each. AdamW is used as the optimizer. Class-weighted Cross Entropy Loss is used as the loss function, with class weights 1.0/1.5/20.0/5.0 for NM/FM/AFM/FiM respectively (this penalizes errors in minority classes more to address unbalance in the dataset). It is found that ALIGNN requires a larger learning rate to converge compared to the other two models. The empirical values of learning rate: 0.001, 0.0001 are assigned to ALIGNN, LLM (including multi-modal).

\section{Results and analysis}
\label{test_result}

\label{GNN algorithm validation}
\subsection{GNN Model Performance Validation}

To evaluate ALIGNN performance on the problem of the paper, two others GNN models, CGCNN v2019 \cite{PhysRevLett.120.145301} and coGN \cite{10.1039/d4dd00018h}, which are top performing models from the Materials Project matbench v0.1 matbench\_mp\_is\_metal task leaderboard \cite{Dunn2020} were tested using a gcnn\_keras implementation \cite{REISER2021100095}. The aforementioned training/validation/test data splits and training configuration were used. As shown in Table~\ref{gnn_result-table}, ALIGNN shows better overall performance (F1) than the other two.

\begin{table}[!ht]
  \caption{Per-class precision, recall and F1 for GNN models On Test Dataset}
  \label{gnn_result-table}
  \centering
  \begin{tabular}{llll}
    \toprule
    Class      & ALIGNN  & CGCNN v2019  &  coGN     \\
    \midrule
    Non-magnetic     & 96\%/88\%/91.8\%        & 93\%/92\%/93\%     & 94\%/93\%/93\%      \\
    \midrule
    Ferromagnetic     & 74\%/83\%/78.2\%       & 79\%/64\%/70\%     & 79\%/61\%/69\%       \\
    \midrule
    Antiferromagnetic    & 32\%/42\%/36.3\%    & 16\%/64\%/26\%    & 16\%/53\%/24\%        \\
    \midrule
    Ferrimagnetic     & 54\%/55\%/54.5\%       & 47\%/59\%/52\%     & 40\%/64\%/49\%        \\
    \bottomrule
  \end{tabular}
\end{table}

\subsection{LLM Model Performance Validation}
\label{LLM algorithm validation}
To evaluate GPT-2's performance, different LLM models used in material science research are evaluated on the dataset of the paper on same training and test setting including LLM-Prop (google T5 v1.1 small encoder)\cite{rubungo2023llmproppredictingphysicalelectronic}, MatBert-base\cite{walker2021impact}, AMD-LLAMA-135M\cite{amd-llama-135m}, distilled GPT-2 \cite{distilled-gpt2} in addition to GPT-2 used in the paper. It can be seen from Table \ref{llm_result-table1} and Table \ref{llm_result-table2} that GPT-2 achieves the best performance. From the model complexity Table \ref{model-complex}, ALIGNN achieves good accuracy relative to its model size: for example, it outperforms LLMProp with much fewer parameters. For edge devices with limited computational capacity, ALIGNN can be a good choice, as it offers a good trade-off between resource requirement and model result accuracy.

\begin{table}[!ht]
  \caption{Per-class precision, recall and F1 for LLM models On Test Dataset}
  \label{llm_result-table1}
  \centering
  \begin{tabular}{llll}
    \toprule
    Class      &   LLMPROP  & distilled GPT-2   \\
    \midrule
    Non-magnetic           & 91.7\%/92\%/92\%    &  93.7\%/92.3\%/93\%   \\
    \midrule
    Ferromagnetic           &  77.8\%/70.8\%/74\%     &  79\%/79\%/79\%      \\
    \midrule
    Antiferromagnetic        & 24\%/34.5\%/28.3\%      &  43.6\%/47.2\%/45.3\%  \\
    \midrule
    Ferrimagnetic          & 43\%/55.3\%/48.3\%       &  51\%/58\%/54.4\%  \\
    \bottomrule
  \end{tabular}
\end{table}

\begin{table}[!ht]
  \caption{Per-class precision, recall and F1 for LLM models On Test Dataset}
  \label{llm_result-table2}
  \centering
  \begin{tabular}{lllll}
    \toprule
    Class      & GPT-2  & MatBert &   AMD-LLAMA    \\
    \midrule
    Non-magnetic     & 94\%/94\%/94\%        & 93\%/94\%/93\%      &  92.7\%/94\%/93.5\%   \\
    \midrule
    Ferromagnetic     & 81\%/82\%/81.5\%       & 80\%/81\%/80\%     &  79.5\%/80\%/79.8\%      \\
    \midrule
    Antiferromagnetic    & 50\%/44\%/46.8\%    & 47\%/41\%/44\%     &  37.6\%/28.3\%/32.3\%  \\
    \midrule
    Ferrimagnetic     & 57\%/52\%/54.4\%       & 59\%/50\%/54\%     &  57.6\%/50.6\%/44\%  \\
    \bottomrule
  \end{tabular}
\end{table}

\subsection{Baseline Models - Performance Validation }
\begin{figure}[!ht]
    \centering
    \includegraphics[width=0.92\textwidth]{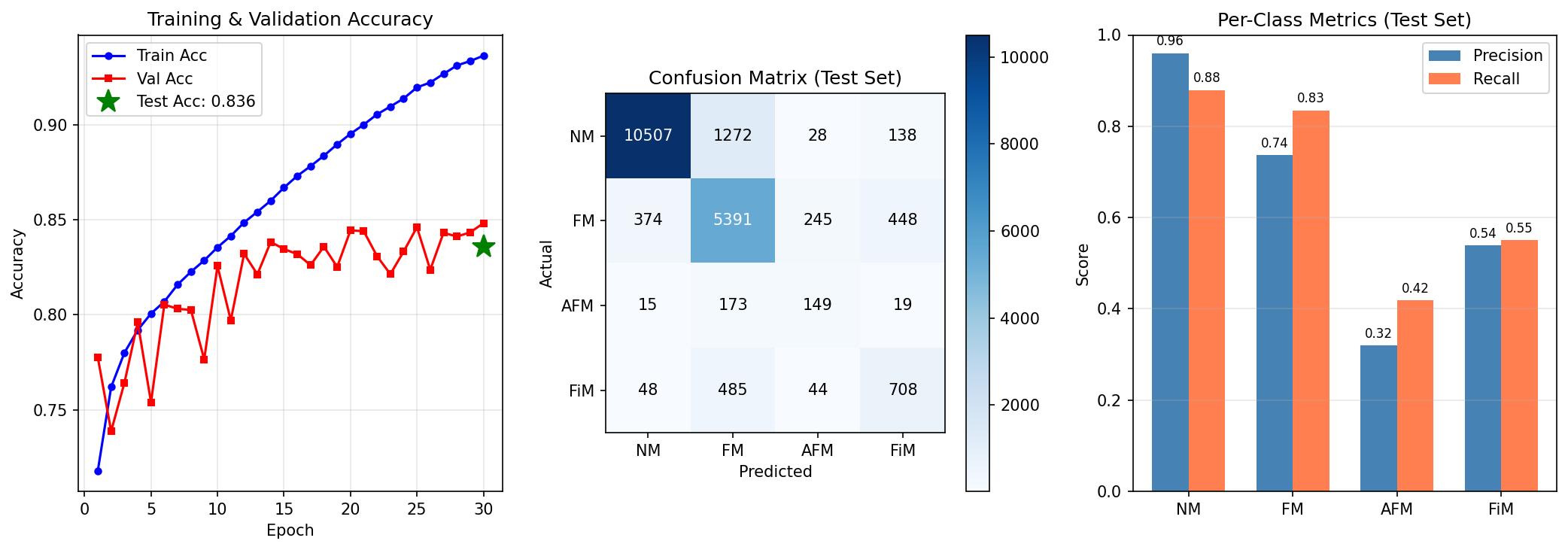}
    \caption{ALIGNN training, validation and testing results}
    \label{fig:gnn_only}
\end{figure}

\begin{figure}[!ht]
    \centering
    \includegraphics[width=0.92\textwidth]{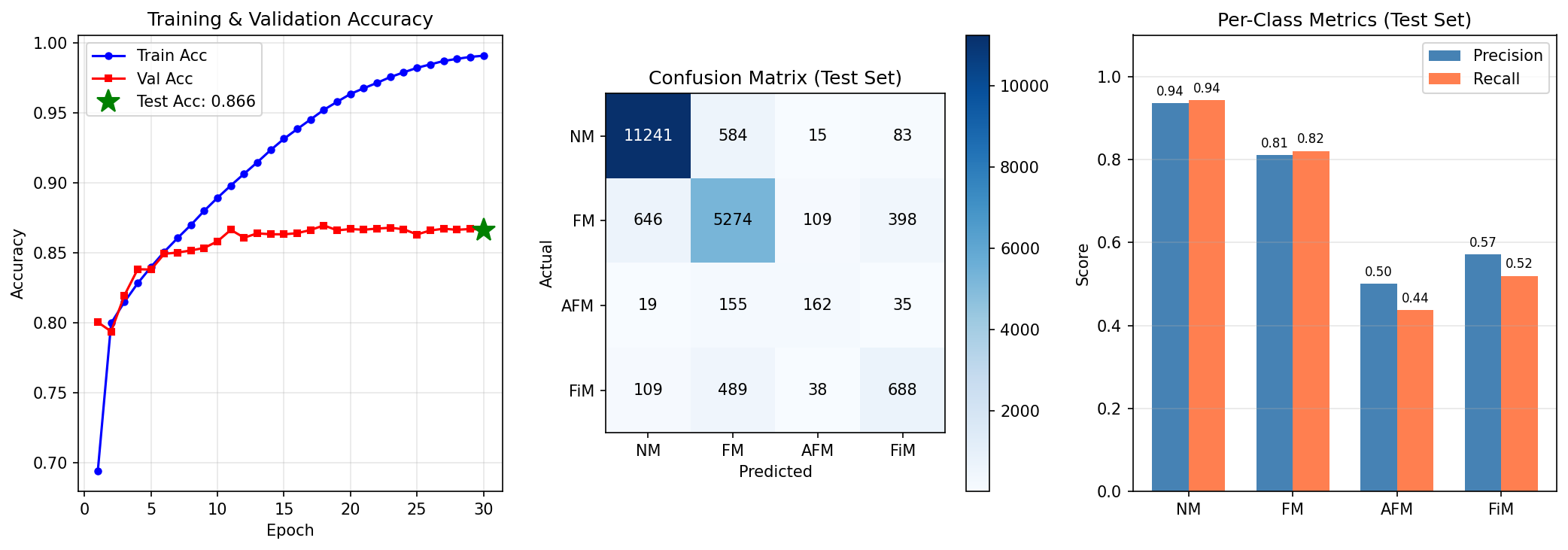}
    \caption{GPT-2 training, validation and testing results}
    \label{fig:llm_only}
\end{figure}

\begin{figure}[!ht]
    \centering
    \includegraphics[width=0.92\textwidth]{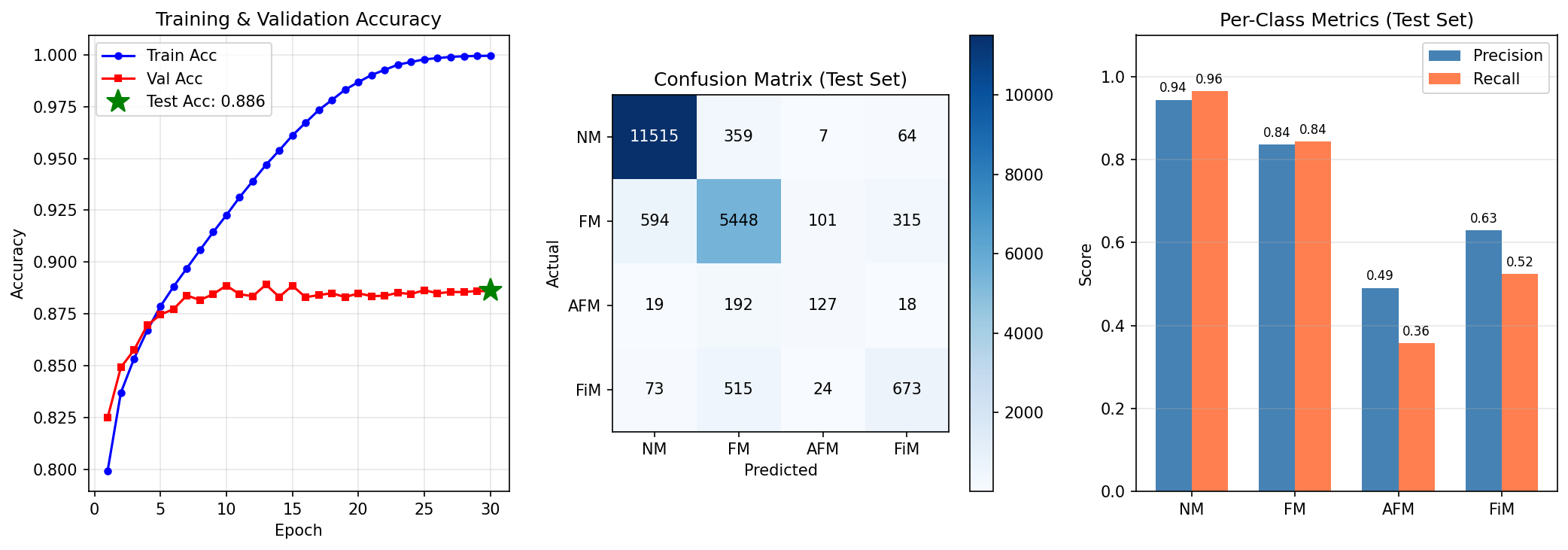}
    \caption{ALIGNN GPT-2 feature fusion multi-modal training, validation and testing results}
    \label{fig:gnn_llm}
\end{figure}

Three models - ALIGNN, GPT-2 base, and intermediate-fusion multi-modal approaches - were selected as the baseline models for comparison. Their training and validation results are presented in Fig.~\ref{fig:gnn_only}, Fig.~\ref{fig:llm_only} and Fig.~\ref{fig:gnn_llm}.  All three approaches converge rapidly: training and validation accuracy quickly exceed 90\% and 80\%, respectively, and plateau at around 30 epochs. Testing results for each model are shown in Fig.~\ref{fig:gnn_only}, Fig.~\ref{fig:llm_only} and Fig.~\ref{fig:gnn_llm} with confusion matrix, precision and recall for each class.


The ALIGNN-LLM multi-modal approach achieves the best overall model accuracy across all four classes, obtaining an accuracy of 88.6\% compared to 83.6\% from ALIGNN alone and 86.6\% from LLM alone.

The per-class testing precision, recall and F1 are shown in Table~\ref{result-table}.
As demonstrated, the ALIGNN + GPT-2 intermediate-fusion multi-modal approach achieves comparable to better classification precision and recall compared to standalone ALIGNN and GPT-2 across most classes (with the exception of the AFM class). Of the four classes, AFM has the lowest classification precision and recall for each model. 

\begin{table}[!ht]
  \caption{Per-class precision, recall and F1 for all three models on test dataset}
  \label{result-table}
  \centering
  \begin{tabular}{llll}
    \toprule
    Class      & ALIGNN only & GPT-2 only &  Intermediate fusion multi-modal    \\
    \midrule
    Non-magnetic     & 96\%/88\%/91.8\%       & 94\%/94\%/94.0\%     & 94.8\%/96.1\%/95.4\%       \\
    \midrule
    Ferromagnetic     & 74\%/83\%/78.2\%     & 81\%/82\%/81.5\%     & 83.3\%/85.2\%/84.2\%       \\
    \midrule
    Antiferromagnetic    & 32\%/42\%/36.3\%   & 50\%/44\%/46.8\%     & 46.9\%/36.5\%/41.1\%        \\
    \midrule
    Ferrimagnetic     & 54\%/55\%/54.5\%     & 57\%/52\%/54.4\%     & 64.2\%/53.4\%/58.3\%        \\
    \bottomrule
  \end{tabular}
\end{table}

Three forms of "randomly guessing" labels are investigated: the accuracy of "Uniform Guessing", "Proportional Guessing" and "Zero Rule" can be calculated as 25\%, 46.13\% and 59.48\% respectively (see Section \ref{guess}). This clearly demonstrates the strength of the models over random guessing.

\subsection{Precision-Recall AUC Analysis}
\begin{figure}[t!]
    \centering
    
    \begin{subfigure}[t]{0.45\textwidth}
        \includegraphics[width=\textwidth]{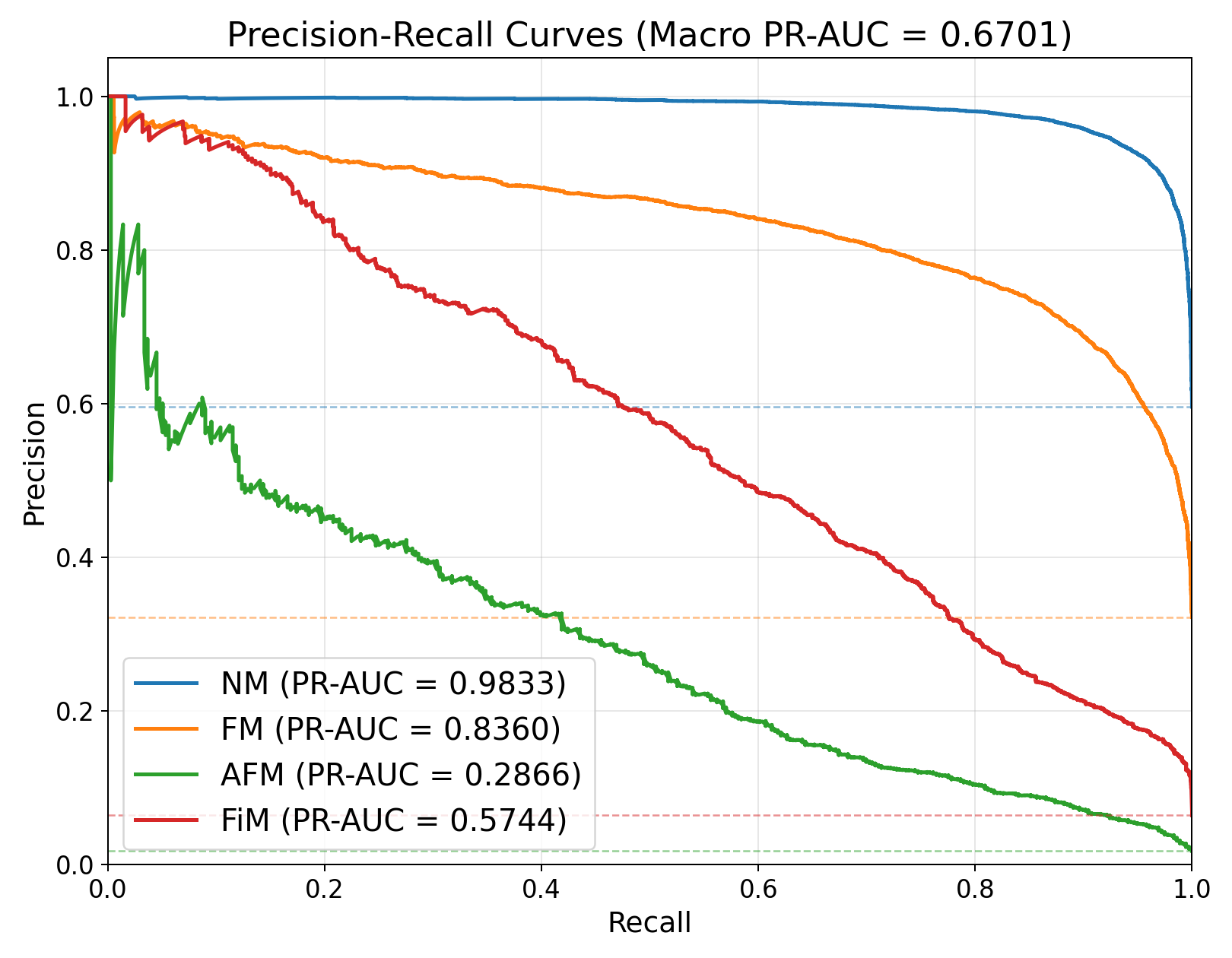}
        \caption{ALIGNN-only PR curve}
        \label{fig:gnn_auc_only}
    \end{subfigure}
    ~
    \begin{subfigure}[t]{0.45\textwidth}
        \includegraphics[width=\textwidth]{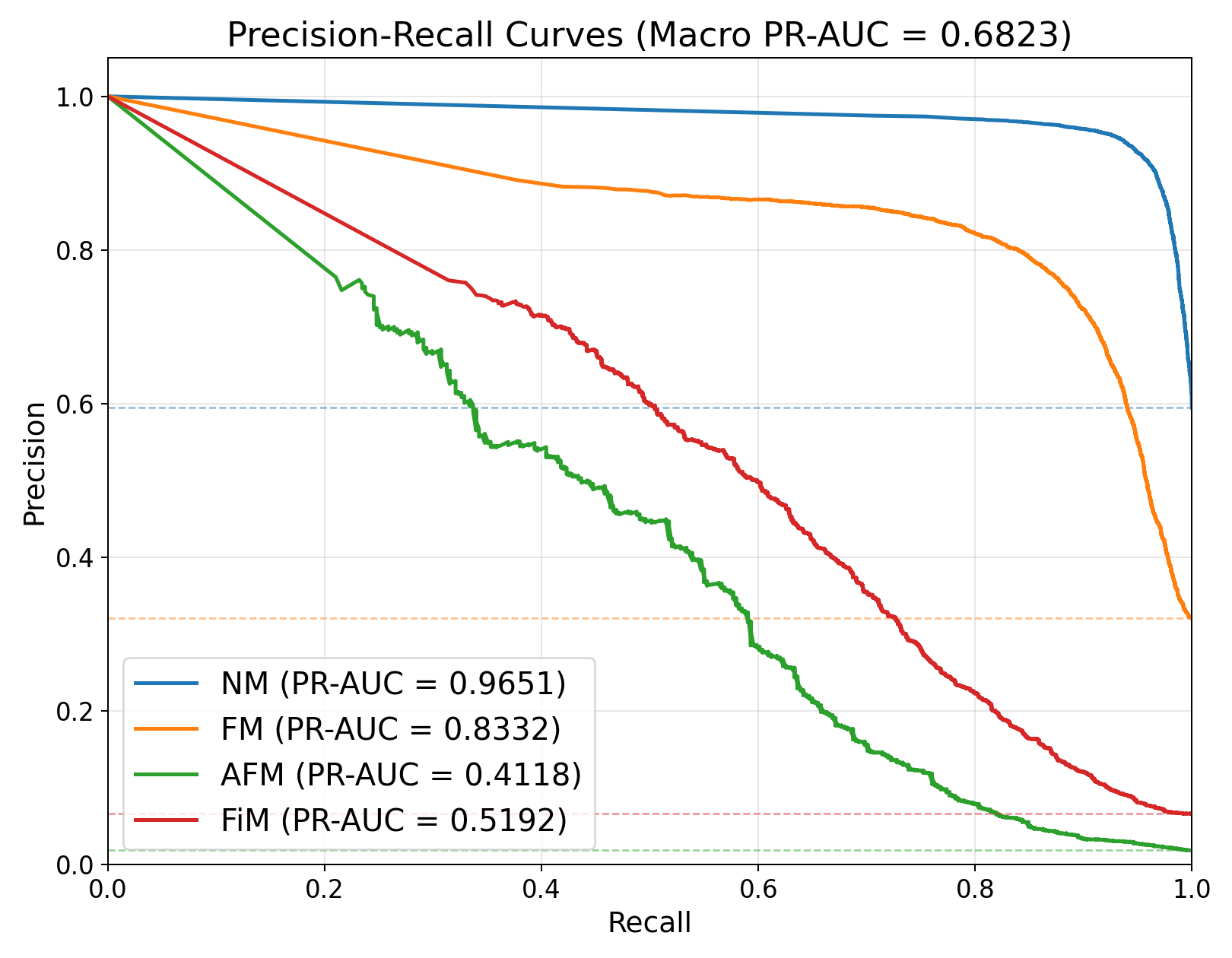}
        \caption{LLM-only PR curve}
        \label{fig:llm_auc_only}
    \end{subfigure}
    ~
    \begin{subfigure}[t]{0.45\textwidth}
        \includegraphics[width=\textwidth]{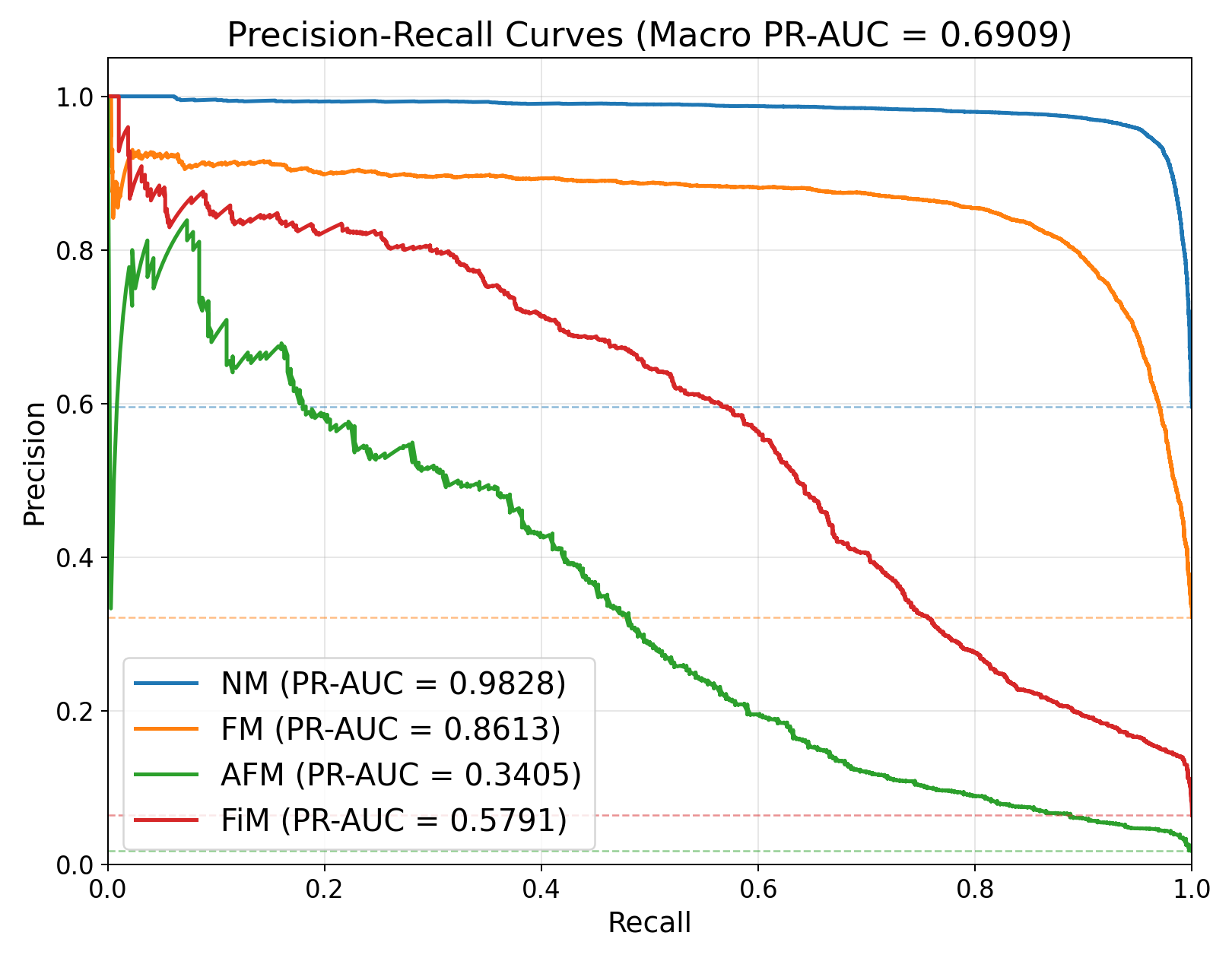}
        \caption{ALIGNN-LLM feature fusion multi-modal PR curve}
        \label{fig:gnn_llm_auc}
    \end{subfigure}

    \caption{PR curves for all three models}
    \label{fig:all_auc}
\end{figure}

To analyze the model performance on test data with unbalanced classes, 
model prediction probability values for each class (i.e. output of the final Softmax layer) are logged during the test process. Afterwards, the precision-recall (PR) curve for each class is plotted by choosing a class (one vs rest) and computing precision and recall for this class while iterating the decision threshold from low to high. The PR curve shows the tradeoff between precision and recall. The curves for each model are shown in Fig. \ref{fig:all_auc}. Subsequently, the Area Under Curve (AUC) is calculated to determine how well the model maintains high precision while increasing recall. A higher AUC is preferred, as having high precision and recall is more ideal than high precision or recall alone. The ALIGNN-LLM multi-modal approach shows better performance in most classes compared to the other two models.

In the multi-modal PR curve (Fig. \ref{fig:gnn_llm_auc}), precision starts off abnormally low for AFM. This is because the decision threshold starts off very close to 1, so the model makes very few positive predictions. The low precision shows that some of the model's most confident predictions are incorrect. This can be improved in future work through further model calibration.

\subsection{Multi-Modal fusion approach performance comparison}
\label{fusion algorithm validation}

For the three fusion approaches, the overall accuracy of the model are 88.77\% for input (early) fusion and 88.6\% for feature (intermediate) fusion and 88.6\% for gated fusion. The per-class precision/recall/F1 are shown in Table~\ref{fusion-result-table}. 

\begin{table}[!ht]
  \caption{Per-class precision, recall and F1 for all three fusion approaches on test dataset}
  \label{fusion-result-table}
  \centering
  \begin{tabular}{llll}
    \toprule
    Class      & Input (early) fusion & Feature (intermediate) fusion &  Gated (late) fusion    \\
    \midrule
    Non-magnetic     & 95\%/96\%/95.5\%       & 94.8\%/96.1\%/95.4\% & 95\%/95.5\%/95.3\%      \\
    \midrule
    Ferromagnetic     & 83.2\%/85.5\%/84.3\%     & 83.3\%/85.2\%/84.2\% &  82.3\%/85.8\%/84\%  \\
    \midrule
    Antiferromagnetic    & 44.3\%/28.3\%/34.6\%   & 46.9\%/36.5\%/41.1\%     &  54.8\%/30.0\%/39\%     \\
    \midrule
    Ferrimagnetic     & 62.7\%/55.6\%/59\%    & 64.2\%/53.4\%/58.3\%     &    62.3\%/54.2\%/58\%   \\
    \bottomrule
  \end{tabular}
\end{table}

\subsection{Model Complexity Analysis}

The complexity of the model is shown in Table~\ref{model-complex} including LLM-Prop (google T5 small v1.1 encoder) and Matbert-base.

\begin{table}[!ht]
  \caption{Model Complexity}
  \label{model-complex}
  \centering
  \begin{tabular}{lllllll}
    \toprule
          &   GPT-2 base & ALIGNN & LLM-Prop & Matbert & AMD-Llama & distilled GPT2 \\
    \midrule
    Parameters     & 124M   & 4.3M & 36M  & 110M & 130M &   88M  \\
 
    \bottomrule
  \end{tabular}
\end{table}

\section{Conclusion and Future Work}

In this paper, a novel ALIGNN-LLM multi-modal approach to predict material magnetic ordering is proposed. The multi-modal design uses ALIGNN to create crystal structure embedding and LLM to create semantic text embedding, and fuses material embedding information together to perform classification. The test results show that this multi-modal approach performs better than ALIGNN-only and LLM-only approaches. For further improvement, the dataset can be balanced by including more samples (especially for the Antiferromagnetic class) when more data becomes available. Moreover, the higher performance of the multi-modal approach comes at the cost of memory usage and computational cost. Nevertheless, the results obtained in this study highlight the potential of the approach proposed herein.

This approach can also be adapted to handle regression problems or multi-properties prediction with the use of more complicated LLM models. Moreover, this approach can be applied to problems of material generation by utilizing the decoding capabilities of the LLM block - the fused output from the LLM and ALIGNN blocks (as detailed above) can be fed into the LLM decoder to generate new materials with desired properties.

\section{Data Availability} The data used in this study are available at the references in \citet{PhysRevMaterials.8.114405}. Datasets generated during the current study are available from the corresponding author on reasonable request.

\begin{appendices}

\section{Random Guessing Algorithms}
\label{guess}

Three "random guessing" algorithms are employed as a baseline in this paper against which the ALIGNN-LLM multi-modal approach is compared: uniform guessing, proportional guessing, and the Zero Rule.

For uniform guessing, the algorithm predicts each of the four classes (NM, FM, AFM, FiM) with equal probability (i.e. 25\% chance to predict each class). The accuracy of this algorithm is 25\%.

For proportional guessing, the algorithm predicts each class with a probability equal to its prevalence in the dataset. For instance, NM samples make up 59.48\% of the dataset, so the algorithm will guess NM with a probability of 59.48\%. This algorithm has an accuracy of 46.13\%:

\begin{align*}
P_{\text{proportional}} &= \sum_{i\in\{ \text{NM, FM, AFM, FiM} \}} P(\text{algorithm guesses class } i)\cdot P(\text{correct label is class } i) \\
&= 59.48\%\cdot 59.48\%+32.06\%\cdot 32.06\% + 1.85\%\cdot 1.85\% + 6.61\%\cdot 6.61\% \\
&= 46.13\%
\end{align*}

The Zero Rule algorithm guesses the class with the largest proportion of the dataset each time. For the dataset used in this study, NM makes up the largest percentage of the data at 59.48\%; hence, the model guesses NM each time. The accuracy of this model is 59.48\%.

\end{appendices}


\begin{thebibliography}{99}
\ifx \bisbn   \undefined \def \bisbn  #1{ISBN #1}\fi
\ifx \binits  \undefined \def \binits#1{#1}\fi
\ifx \bauthor  \undefined \def \bauthor#1{#1}\fi
\ifx \batitle  \undefined \def \batitle#1{#1}\fi
\ifx \bjtitle  \undefined \def \bjtitle#1{#1}\fi
\ifx \bvolume  \undefined \def \bvolume#1{\textbf{#1}}\fi
\ifx \byear  \undefined \def \byear#1{#1}\fi
\ifx \bissue  \undefined \def \bissue#1{#1}\fi
\ifx \bfpage  \undefined \def \bfpage#1{#1}\fi
\ifx \blpage  \undefined \def \blpage #1{#1}\fi
\ifx \burl  \undefined \def \burl#1{\textsf{#1}}\fi
\ifx \doiurl  \undefined \def \doiurl#1{\url{https://doi.org/#1}}\fi
\ifx \betal  \undefined \def \betal{\textit{et al.}}\fi
\ifx \binstitute  \undefined \def \binstitute#1{#1}\fi
\ifx \binstitutionaled  \undefined \def \binstitutionaled#1{#1}\fi
\ifx \bctitle  \undefined \def \bctitle#1{#1}\fi
\ifx \beditor  \undefined \def \beditor#1{#1}\fi
\ifx \bpublisher  \undefined \def \bpublisher#1{#1}\fi
\ifx \bbtitle  \undefined \def \bbtitle#1{#1}\fi
\ifx \bedition  \undefined \def \bedition#1{#1}\fi
\ifx \bseriesno  \undefined \def \bseriesno#1{#1}\fi
\ifx \blocation  \undefined \def \blocation#1{#1}\fi
\ifx \bsertitle  \undefined \def \bsertitle#1{#1}\fi
\ifx \bsnm \undefined \def \bsnm#1{#1}\fi
\ifx \bsuffix \undefined \def \bsuffix#1{#1}\fi
\ifx \bparticle \undefined \def \bparticle#1{#1}\fi
\ifx \barticle \undefined \def \barticle#1{#1}\fi
\bibcommenthead
\ifx \bconfdate \undefined \def \bconfdate #1{#1}\fi
\ifx \botherref \undefined \def \botherref #1{#1}\fi
\ifx \bchapter \undefined \def \bchapter#1{#1}\fi
\ifx \bbook \undefined \def \bbook#1{#1}\fi
\ifx \bcomment \undefined \def \bcomment#1{#1}\fi
\ifx \oauthor \undefined \def \oauthor#1{#1}\fi
\ifx \citeauthoryear \undefined \def \citeauthoryear#1{#1}\fi
\ifx \endbibitem  \undefined \def \endbibitem {}\fi
\ifx \bconflocation  \undefined \def \bconflocation#1{#1}\fi
\ifx \arxivurl  \undefined \def \arxivurl#1{\textsf{#1}}\fi


\bibitem[\protect\citeauthoryear{Parr and Yang}{1989}]{dft2}
\begin{bbook}
\bauthor{\bsnm{Parr}, \binits{R.G.}},
\bauthor{\bsnm{Yang}, \binits{W.}}:
\bbtitle{Density-Functional Theory of Atoms and Molecules}.
\bpublisher{Oxford University Press}, \blocation{Oxford}
(\byear{1989})
\doiurl{10.1093/oso/9780195092769.001.0001}
\end{bbook}
\endbibitem

\bibitem[\protect\citeauthoryear{March}{1999}]{dft3}
\begin{bbook}
\bauthor{\bsnm{March}, \binits{N.H.}}:
\bbtitle{Electron Correlation in the Solid State}.
\bpublisher{ICP}, \blocation{London}
(\byear{1999})
\doiurl{10.1142/p174}
\end{bbook}
\endbibitem

\bibitem[\protect\citeauthoryear{Callaway and March}{1984}]{dft1}
\begin{barticle}
\bauthor{\bsnm{Callaway}, \binits{J.}},
\bauthor{\bsnm{March}, \binits{N.H.}}:
\batitle{Density Functional Methods: Theory and Applications.}
\bjtitle{Solid State Physics}
\bvolume{38},
\bfpage{135}--\blpage{221}
(\byear{1984})
\doiurl{10.1016/S0081-1947(08)60313-6}
\end{barticle}
\endbibitem

\bibitem[\protect\citeauthoryear{Sch\"{u}tt et~al.}{2017}]{NIPS2017_303ed4c6}
\begin{bchapter}
\bauthor{\bsnm{Sch\"{u}tt}, \binits{K.}},
\bauthor{\bsnm{Kindermans}, \binits{P.-J.}},
\bauthor{\bsnm{Sauceda Felix}, \binits{H.E.}},
\bauthor{\bsnm{Chmiela}, \binits{S.}},
\bauthor{\bsnm{Tkatchenko}, \binits{A.}},
\bauthor{\bsnm{M\"{u}ller}, \binits{K.-R.}}:
\bctitle{\textit{SchNet: A continuous-filter convolutional neural network for modeling quantum interactions}}.
In: \bbtitle{Advances in Neural Information Processing Systems}
\beditor{\bsnm{Guyon}, \binits{I.}},
\beditor{\bsnm{Luxburg}, \binits{U.V.}},
\beditor{\bsnm{Bengio}, \binits{S.}},
\beditor{\bsnm{Wallach}, \binits{H.}},
\beditor{\bsnm{Fergus}, \binits{R.}},
\beditor{\bsnm{Vishwanathan}, \binits{S.}},
\beditor{\bsnm{Garnett}, \binits{R.}}
\bvolume{30}
\bpublisher{Curran Associates, Inc.}
(\byear{2017})
\url{https://proceedings.neurips.cc/paper_files/paper/2017/file/303ed4c69846ab36c2904d3ba8573050-Paper.pdf}
\end{bchapter}
\endbibitem

\bibitem[\protect\citeauthoryear{Xie and Grossman}{2018}]{PhysRevLett.120.145301}
\begin{barticle}
\bauthor{\bsnm{Xie}, \binits{T.}},
\bauthor{\bsnm{Grossman}, \binits{J.C.}}:
\batitle{\textit{Crystal Graph Convolutional Neural Networks for an Accurate and Interpretable Prediction of Material Properties}}.
\bjtitle{Phys. Rev. Lett.}
\bvolume{120}(\bissue{14}),
\bfpage{145301}
(\byear{2018})
\doiurl{10.1103/PhysRevLett.120.145301}
\end{barticle}
\endbibitem

\bibitem[\protect\citeauthoryear{Louis et~al.}{2020}]{D0CP01474E}
\begin{barticle}
\bauthor{\bsnm{Louis}, \binits{S.-Y.}},
\bauthor{\bsnm{Zhao}, \binits{Y.}},
\bauthor{\bsnm{Nasiri}, \binits{A.}},
\bauthor{\bsnm{Wang}, \binits{X.}},
\bauthor{\bsnm{Song}, \binits{Y.}},
\bauthor{\bsnm{Liu}, \binits{F.}},
\bauthor{\bsnm{Hu}, \binits{J.}}:
\batitle{\textit{Graph convolutional neural networks with global attention for improved materials property prediction}}.
\bjtitle{Phys. Chem. Chem. Phys.}
\bvolume{22}(\bissue{32}),
\bfpage{18141}--\blpage{18148}
(\byear{2020})
\doiurl{10.1039/D0CP01474E}
\end{barticle}
\endbibitem

\bibitem[\protect\citeauthoryear{Choudhary and DeCost}{2021}]{Choudhary}
\begin{barticle}
\bauthor{\bsnm{Choudhary}, \binits{K.}},
\bauthor{\bsnm{DeCost}, \binits{B.}}:
\batitle{\textit{Atomistic Line Graph Neural Network for improved materials property predictions}}.
\bjtitle{npj Computational Materials 7, 185}
(\byear{2021})
\doiurl{10.1038/s41524-021-00650-1}
\end{barticle}
\endbibitem



\bibitem[\protect\citeauthoryear{Vaswani et~al.}{2017}]{NIPS2017_3f5ee243}
\begin{bchapter}
\bauthor{\bsnm{Vaswani}, \binits{A.}},
\bauthor{\bsnm{Shazeer}, \binits{N.}},
\bauthor{\bsnm{Parmar}, \binits{N.}},
\bauthor{\bsnm{Uszkoreit}, \binits{J.}},
\bauthor{\bsnm{Jones}, \binits{L.}},
\bauthor{\bsnm{Gomez}, \binits{A.N.}},
\bauthor{\bsnm{Kaiser}, \binits{{\L}.}},
\bauthor{\bsnm{Polosukhin}, \binits{I.}}:
\bctitle{\textit{Attention is All you Need}}.
In: \bbtitle{Advances in Neural Information Processing Systems}
\beditor{\bsnm{Guyon}, \binits{I.}},
\beditor{\bsnm{Luxburg}, \binits{U.V.}},
\beditor{\bsnm{Bengio}, \binits{S.}},
\beditor{\bsnm{Wallach}, \binits{H.}},
\beditor{\bsnm{Fergus}, \binits{R.}},
\beditor{\bsnm{Vishwanathan}, \binits{S.}},
\beditor{\bsnm{Garnett}, \binits{R.}}
\bvolume{30}
\bpublisher{Curran Associates, Inc.}
(\byear{2017})
\url{https://proceedings.neurips.cc/paper_files/paper/2017/file/3f5ee243547dee91fbd053c1c4a845aa-Paper.pdf}
\end{bchapter}
\endbibitem

\bibitem[\protect\citeauthoryear{Yan et~al.}{2024}]{yan2024complete}
\begin{bchapter}
\bauthor{\bsnm{Yan}, \binits{K.}},
\bauthor{\bsnm{Fu}, \binits{C.}},
\bauthor{\bsnm{Qian}, \binits{X.}},
\bauthor{\bsnm{Qian}, \binits{X.}},
\bauthor{\bsnm{Ji}, \binits{S.}}:
\bctitle{\textit{Complete and Efficient Graph Transformers for Crystal Material Property Prediction}}.
In: \bbtitle{The Twelfth International Conference on Learning Representations}
(\byear{2024})
\url{https://openreview.net/forum?id=BnQY9XiRAS}
\end{bchapter}
\endbibitem

\bibitem[\protect\citeauthoryear{Walker et~al.}{2021}]{walker2021impact}
\begin{barticle}
\bauthor{\bsnm{Walker}, \binits{N.}},
\bauthor{\bsnm{Trewartha}, \binits{A.}},
\bauthor{\bsnm{Huo}, \binits{H.}},
\bauthor{\bsnm{Lee}, \binits{S.}},
\bauthor{\bsnm{Cruse}, \binits{K.}},
\bauthor{\bsnm{Dagdelen}, \binits{J.}},
\bauthor{\bsnm{Dunn}, \binits{A.}},
\bauthor{\bsnm{Persson}, \binits{K.}},
\bauthor{\bsnm{Ceder}, \binits{G.}},
\bauthor{\bsnm{Jain}, \binits{A.}}:
\batitle{\textit{The Impact of Domain-Specific Pre-Training on Named Entity Recognition Tasks in Materials Science}}.
\bjtitle{Available at SSRN 3950755}
(\byear{2021})
\end{barticle}
\endbibitem

\bibitem[\protect\citeauthoryear{Rubungo et~al.}{2023}]{rubungo2023llmproppredictingphysicalelectronic}
\begin{botherref}
\oauthor{\bsnm{Rubungo}, \binits{A.N.}},
\oauthor{\bsnm{Arnold}, \binits{C.}},
\oauthor{\bsnm{Rand}, \binits{B.P.}},
\oauthor{\bsnm{Dieng}, \binits{A.B.}}:
LLM-Prop: Predicting Physical And Electronic Properties Of Crystalline Solids From Their Text Descriptions.
arXiv preprint arXiv:2310.14029
\url{https://arxiv.org/abs/2310.14029}
(\byear{2023})
\end{botherref}
\endbibitem

\bibitem[\protect\citeauthoryear{Li et~al.}{2025}]{li2025hybrid}
\begin{barticle}
\bauthor{\bsnm{Li}, \binits{Y.}},
\bauthor{\bsnm{Gupta}, \binits{V.}},
\bauthor{\bsnm{Kilic}, \binits{M.N.T.}},
\bauthor{\bsnm{Choudhary}, \binits{K.}},
\bauthor{\bsnm{Wines}, \binits{D.}},
\bauthor{\bsnm{Liao}, \binits{W.-K.}},
\bauthor{\bsnm{Choudhary}, \binits{A.}},
\bauthor{\bsnm{Agrawal}, \binits{A.}}:
\batitle{\textit{Hybrid-LLM-GNN: integrating large language models and graph neural networks for enhanced materials property prediction}}.
\bjtitle{Digital Discovery}
(\byear{2025})
\end{barticle}
\endbibitem

\bibitem[\protect\citeauthoryear{Broyles et~al.}{2024}]{PhysRevMaterials.8.114405}
\begin{barticle}
\bauthor{\bsnm{Broyles}, \binits{C.}},
\bauthor{\bsnm{Charles}, \binits{W.}},
\bauthor{\bsnm{Ran}, \binits{S.}}:
\batitle{\textit{Structure-driven prediction of magnetic order in uranium compounds}}.
\bjtitle{Phys. Rev. Mater.}
\bvolume{8}(\bissue{11}),
\bfpage{114405}
(\byear{2024})
\doiurl{10.1103/PhysRevMaterials.8.114405}
\end{barticle}
\endbibitem

\bibitem[\protect\citeauthoryear{gpt2 model}{n.d.}]{gpt2}
\begin{botherref}
gpt2 model.
\url{https://huggingface.co/openai-community/gpt2}
\end{botherref}
\endbibitem

\bibitem[\protect\citeauthoryear{jarvis.core.atoms}{n.d.}]{javisatoms}
\begin{botherref}
jarvis.core.atoms.
\url{https://jarvis-tools.readthedocs.io/en/master/autoapi/jarvis/core/atoms/index.html}
\end{botherref}
\endbibitem

\bibitem[\protect\citeauthoryear{Material Project}{n.d.}]{mproj}
\begin{botherref}
Material Project.
\url{https://next-gen.materialsproject.org/materials}
\end{botherref}
\endbibitem

\bibitem[\protect\citeauthoryear{ALIGNN-BERT}{n.d.}]{chemnlp}
\begin{botherref}
ALIGNN-BERT.
\url{https://github.com/Jonathanlyj/ALIGNN-BERT-TL-crystal}
\end{botherref}
\endbibitem

\bibitem[\protect\citeauthoryear{magndata cif}{n.d.}]{magndata}
\begin{botherref}
magndata cif.
\url{https://www.cryst.ehu.es/magndata/}
\end{botherref}
\endbibitem



\bibitem[\protect\citeauthoryear{Ruff et~al.}{2024}]{10.1039/d4dd00018h}
\begin{barticle}
\bauthor{\bsnm{Ruff}, \binits{R.}},
\bauthor{\bsnm{Reiser}, \binits{P.}},
\bauthor{\bsnm{St{\"u}hmer}, \binits{J.}},
\bauthor{\bsnm{Friederich}, \binits{P.}}:
\batitle{\textit{Connectivity optimized nested line graph networks for crystal structures}}.
\bjtitle{Digital Discovery 3(3), 594--601}
(\byear{2024})
\doiurl{10.1039/d4dd00018h}
\end{barticle}
\endbibitem

\bibitem[\protect\citeauthoryear{Dunn et~al.}{2020}]{Dunn2020}
\begin{barticle}
\bauthor{\bsnm{Dunn}, \binits{A.}},
\bauthor{\bsnm{Wang}, \binits{Q.}},
\bauthor{\bsnm{Ganose}, \binits{A.}},
\bauthor{\bsnm{Dopp}, \binits{D.}},
\bauthor{\bsnm{Jain}, \binits{A.}}:
\batitle{\textit{Benchmarking materials property prediction methods: the Matbench test set and Automatminer reference algorithm}}.
\bjtitle{npj Computational Materials 6, 138}
(\byear{2020})
\doiurl{10.1038/s41524-020-00406-3}
\end{barticle}
\endbibitem

\bibitem[\protect\citeauthoryear{Reiser et~al.}{2021}]{REISER2021100095}
\begin{barticle}
\bauthor{\bsnm{Reiser}, \binits{P.}},
\bauthor{\bsnm{Eberhard}, \binits{A.}},
\bauthor{\bsnm{Friederich}, \binits{P.}}:
\batitle{\textit{Graph neural networks in TensorFlow-Keras with RaggedTensor representation (kgcnn)}}.
\bjtitle{Software Impacts, 100095}
(\byear{2021})
\doiurl{10.1016/j.simpa.2021.100095}
\end{barticle}
\endbibitem

\bibitem[\protect\citeauthoryear{AMD-Llama-135M}{n.d.}]{amd-llama-135m}
\begin{botherref}
AMD-Llama-135M.
\url{https://huggingface.co/distilbert/distilgpt2}
\end{botherref}
\endbibitem

\bibitem[\protect\citeauthoryear{Distilled-GPT2}{n.d.}]{distilled-gpt2}
\begin{botherref}
Distilled-GPT2.
\url{https://huggingface.co/amd/AMD-Llama-135m}
\end{botherref}
\endbibitem

\bibitem[\protect\citeauthoryear{Wan et~al.}{2024}]{wan2024tokensmaterialsleveraginglanguage}
\begin{botherref}
\oauthor{\bsnm{Wan}, \binits{Y.}},
\oauthor{\bsnm{Xie}, \binits{T.}},
\oauthor{\bsnm{Wu}, \binits{N.}},
\oauthor{\bsnm{Zhang}, \binits{W.}},
\oauthor{\bsnm{Kit}, \binits{C.}},
\oauthor{\bsnm{Hoex}, \binits{B.}}:
From Tokens to Materials: Leveraging Language Models for Scientific Discovery.
arXiv preprint arXiv:2410.16165
\url{https://arxiv.org/abs/2410.16165}
(\byear{2024})
\end{botherref}
\endbibitem





\bibitem[\protect\citeauthoryear{llama model}{n.d.}]{meta-llama}
\begin{botherref}
llama model.
\url{https://huggingface.co/meta-llama}
\end{botherref}
\endbibitem




\end{thebibliography}
\end{document}